\documentclass{elsart}
\usepackage{graphicx}
\usepackage{epsfig}

\def\gsim{\;\raise0.3ex\hbox{$>$\kern-0.75em\raise-1.1ex\hbox{$\sim$}}\;}
\def\lsim{\;\raise0.3ex\hbox{$<$\kern-0.75em\raise-1.1ex\hbox{$\sim$}}\;}

\begin{document}

\begin{frontmatter}
{\hfill \small DSF-17/2005, IFIC/05-36}

\title{The aperture for UHE tau neutrinos\\
of the Auger fluorescence detector\\
using a Digital Elevation Map}
\author[Napoli]{Gennaro Miele},
\author[IFIC]{Sergio Pastor},
\author[Napoli]{Ofelia Pisanti}
\address[Napoli]{Dipartimento di Scienze Fisiche, Universit\`{a} di Napoli
``Federico II'' and INFN Sezione di Napoli, Complesso Universitario di
Monte S.\ Angelo, Via Cinthia, I-80126 Napoli, Italy}
\address[IFIC]{Instituto de F\'{\i}sica Corpuscular (CSIC-Universitat de
Val\`{e}ncia),\\ Ed.\ Institutos de Investigaci\'{o}n, Apdo.\ 22085,
E-46071 Valencia, Spain}

\begin{abstract}
We perform a new study of the chances of the fluorescence detector
(FD) at the Pierre Auger Observatory to detect the tau leptons
produced by Earth-skimming ultra high energy $\nu_\tau$'s. We present
a new and more detailed evaluation of the effective aperture of the FD
that considers a reliable fiducial volume for the experimental set
up. In addition, we take into account the real elevation profile of
the area near Auger. We find a significant increase in the number of
expected events with respect to the predictions of a previous
semi-analytical determination, and our results show the enhancement
effect for neutrino detection from the presence of the near mountains.
\end{abstract}


\end{frontmatter}

Neutrinos constitute one of the components of the cosmic radiation in
the ultra high energy (UHE) regime. Since we have detected ultra high
energy cosmic rays (UHECR), the presence of a secondary UHE neutrino
flux is guaranteed as a result of the $\pi$-photo\-production, due to
the interaction of hadronic UHECR with the cosmic microwave
background.  The detection of these {\it cosmogenic neutrinos}
\cite{cosmogenic,cosmogenic2}, in addition to a possible primary
neutrino flux, would provide precious information on the physics and
position of their powerful astrophysical sources.  On the other hand,
copious neutrino fluxes are also predicted in more exotic {\it
top-down} scenarios where relic massive particles, produced at the
first moments of the Universe, decay into UHE lighter particles, among
which neutrinos and photons are expected. In any case, the detection
of UHE neutrinos would significantly contribute to unveiling the still
unknown origin of UHECR.

Due to the very low expected fluxes and the small neutrino-nucleon
cross section, neutrinos with energies of the order of $10^{18}$ eV
and larger are hardly detectable even in the new generation of giant
array detectors for cosmic radiation, like the Pierre Auger
Observatory (Auger, in short) \cite{Auger,Abraham:2004dt}. The
detection of UHE neutrinos inducing inclined air showers was recently
reviewed in \cite{Zas:2005zz}.  In particular, a promising strategy
concerning the detection of the tau leptons produced by Earth-skimming
UHE $\nu_\tau$'s has been analyzed in a series of papers
\cite{Capelle:1998zz}--\cite{Fargion:2005fa}. UHE $\tau$'s, with
energies in the range $10^{18-21}$ eV, have a decay length not much
larger than the corresponding interaction range. Thus, if a UHE
$\nu_\tau$ crosses the Earth almost horizontally (Earth-skimming) and
interacts in the rock, the produced $\tau$ has a chance to emerge from
the surface and decay in the atmosphere, producing a shower that in
principle can be detected as an up-going or almost horizontal event.

The aim of this letter is to perform a new, more refined, estimate
of the 
effective aperture of the fluorescence detector (FD) at Auger to
Earth-skimming UHE $\nu_\tau$'s. A calculation of the number of
possible up-going $\tau$ showers detectable with the FD
has already been performed in ref.\ \cite{Aramo:2004pr} by using a
semi-analytical computation. Our analysis represents a
considerable improvement with respect to the estimate of this last
work, since it uses a different method for calculating the number
of $\nu_\tau$/$\tau$ events, which now includes a class of tracks
neglected in the previous calculation. 

\begin{figure}[t]
\begin{center}
\includegraphics[width=.95\textwidth]{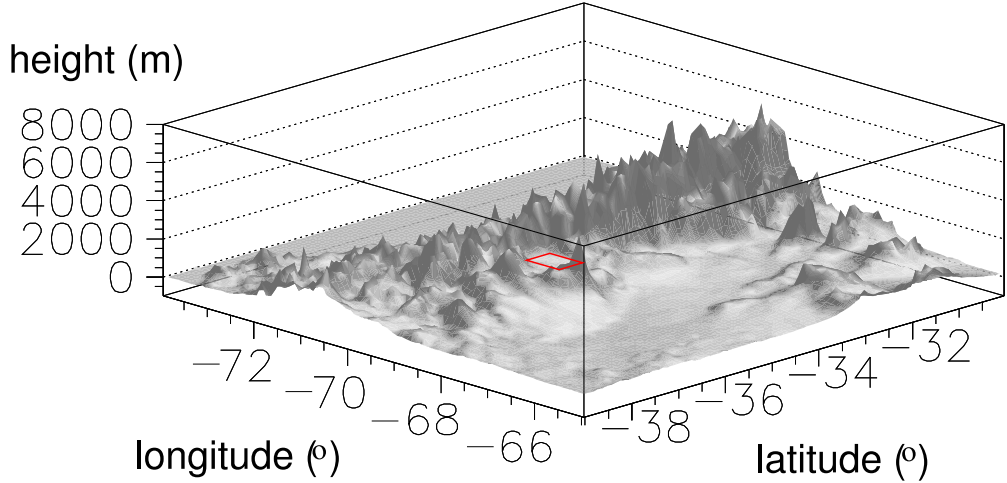}
\caption{\label{mountain3D} A 3D map in longitude and latitude of the
area around Auger with the elevation (not to scale) expressed in
meters. The Auger position and surface, approximated to a rectangle, is
indicated in red.}
\end{center}
\end{figure}

An additional improvement in our analysis comes from considering the
effects of the topology around the Auger observatory site, by using a
Digital Elevation Map (DEM) of the area around the experiment.
A detailed DEM of the Earth surface is provided by ASTER (Advanced
Spaceborne Thermal Emission and Reflection Radiometer) \cite{Aster}
which is an imaging instrument that is flying on Terra, a satellite
launched in December 1999 as part of NASA's Earth Observing
System. The available elevation map, GTOPO30, is a global digital
model where the elevations are regularly spaced at 30-arc seconds.
We show in Fig.\ \ref{mountain3D} a 3D map of the
relevant region around Auger. We will use this DEM to produce a
realistic and statistically significant sample of possible
$\nu_\tau$/$\tau$ tracks crossing the fiducial volume of Auger, that
will be used later to evaluate the real aperture of FD at Auger.

\begin{figure}[t]
\begin{center}
\includegraphics[width=.95\textwidth]{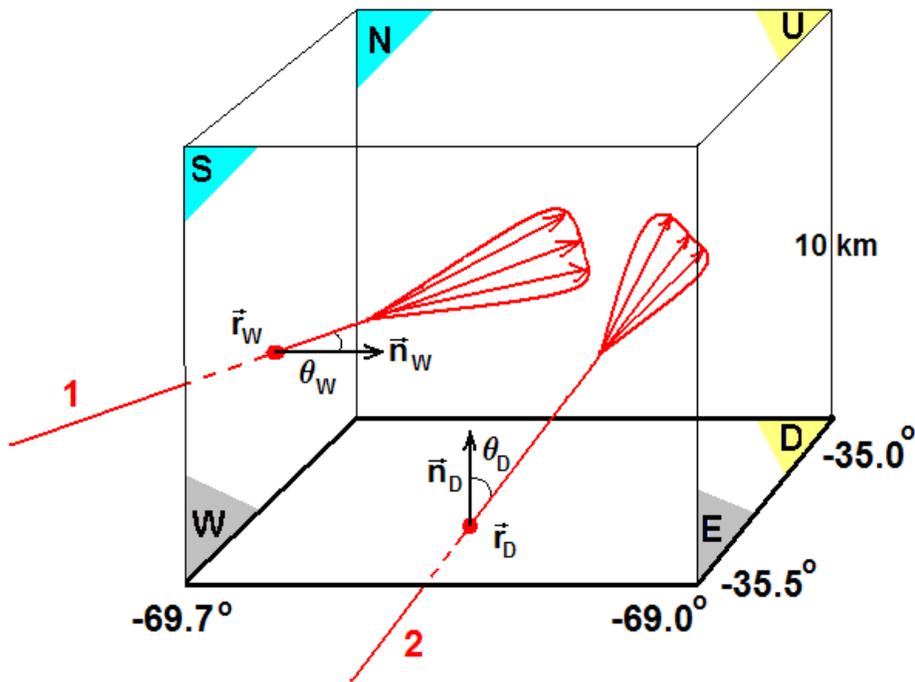}
\caption{A simplified scheme
of the Auger fiducial volume is represented (height not to
scale). The lateral surfaces are labelled by their orientation. Two
examples of entering tracks are also shown.} \label{volume}
\end{center}
\end{figure}

We will define the Auger {\it fiducial} volume as that limited by the
the six lateral surfaces $\Sigma_a$ (the subindex $a=W$, $E$, $N$,
$S$, $U$ and $D$ labels each surface through its orientation: West,
East, North, South, Up, and Down), and with $\Omega_a \equiv
(\theta_a, \phi_a)$ a generic direction of a track entering
$\Sigma_a$, as shown in Fig.\ \ref{volume}. We have considered a
simplified Auger area, given by a $50\times 60$ km rectangle (an
approximation to the real one, see ref.\ \cite{Abraham:2004dt}), while
the height of the fiducial volume was fixed to 10 km in order to be
within the range of detection of the FD eyes. This is a conservative
estimate, since we expect that the effective fiducial volume for the
detection at the FD will be larger.

Let $\Phi_\nu$ be an isotropic flux of $\nu_\tau +
\overline{\nu}_\tau$. By generalizing the formalism developed in ref.\
\cite{Feng:2001ue}, the number of $\tau$ leptons emerging from the
Earth surface with energy $E_\tau$, going through $\Sigma_a$ and
showering in the fiducial volume per unit of time (thus potentially
detectable by the FD), is given by
\begin{eqnarray}
\left( \frac{d N_\tau}{dt} \right)_a &=& D \, \int d\Omega_a \int
dS_a \, \int dE_\nu \, \,
\frac{d\Phi_\nu(E_\nu)}{dE_\nu\,d\Omega_a} \, \nonumber \\
&&\times \int dE_\tau~ \epsilon(E_\tau) \,
\cos\left(\theta_a\right) \,
k_a(E_\nu\,,E_\tau\,;\vec{r}_a\,,\Omega_a)\,\,\,, \label{eq:1}
\end{eqnarray}
where $D$ is the duty cycle and $\epsilon(E_\tau)$ is the
detection efficiency of the FD, respectively \cite{Auger}. The
minimum energy for the $\tau$ leptons, $10^{18}$ eV, is chosen
taking into account the energy threshold for the flourescence
process \cite{Aramo:2004pr}. The kernel
$k_a(E_\nu\,,E_\tau\,;\vec{r}_a\,,\Omega_a)$ is the probability
that an incoming neutrino crossing the Earth with energy $E_\nu$
and direction $\Omega_a$, produces a lepton emerging with energy
$E_\tau$, which enters the fiducial volume through the lateral
surface $dS_a$ at the position $\vec{r}_a$ and decays inside this
volume (see Fig.\ \ref{volume} for the angle definition). In Eq.\
(\ref{eq:1}), due to the very high energy of $\nu_\tau$, we can
assume that in the process $\nu_\tau \, + \, N \rightarrow \tau \,
+ \, X$ the charged lepton is produced along the neutrino
direction.

As already shown in details in ref.\ \cite{Aramo:2004pr}, this
process can occur if the following conditions are
fulfilled,
\begin{itemize}
\item[1)] the $\nu_\tau$ with energy $E_\nu$ has to survive along a
distance $z$ through the Earth;
\item[2)] the neutrino converts into a $\tau$ in the
interval $z, z+dz$;
\item[3)] the created $\tau$ emerges from the Earth before decaying with
energy $E_\tau$;
\item[4)] the $\tau$ lepton enters the fiducial volume through the
lateral surface $\Sigma_a$ at the point $\vec{r}_a$ and decays inside
this volume.
\end{itemize}

\noindent 1) The probability $P_1$ that a neutrino with energy
$E_\nu$ crossing the Earth survives up to a certain distance $z$
inside the rock is
\begin{equation}
P_1=\exp\left\{-\frac{z}{\lambda_{CC}^\nu(E_\nu)}\right\}\,\,\,,
\label{eq:2}
\end{equation}
where
\begin{equation}
\lambda_{CC}^\nu(E_\nu)= \frac{1}{\sigma_{CC}^{\nu
N}(E_\nu)\,\varrho_s\,N_A} \label{eq:3}
\end{equation}
is the charged current (CC) interaction length in rock ($\varrho_s
\simeq 2.65$ g/cm$^{3}$). A detailed discussion of an updated
evaluation of the neutrino-nucleon cross section, $\sigma_{CC}^{\nu
N}(E_\nu)$, can be found in ref.\ \cite{Aramo:2004pr}. The dependence
of $\lambda_{CC}^\nu$ on the particular track direction can be safely
neglected, since it is well known that the interesting events are
almost horizontal and thus the experienced Earth density is
essentially\footnote{We neglect the fact that UHE neutrinos may
traverse a significant amount of water (as those from the West), which
could enhance the arrival of UHE tau neutrinos at the largest
energies, see e.g.\ \cite{Fargion:2003kn}.}  equal to $\varrho_s$.

\noindent 2) The probability for $\nu_\tau \rightarrow \tau$
conversion in the interval $[z,\,z+dz]$ is
\begin{equation}
P_2 \, dz= \frac{dz}{\lambda_{CC}^\nu\,(E_\nu)}\,\,\, .
\label{eq:4}
\end{equation}

\noindent 3) The probability $P_c$ that a charged lepton survives
as it loses its energy travelling through the Earth is described
by the coupled differential equations
\begin{eqnarray}
\frac{dP_c}{dz}&=&-\frac{m_\tau}{c \, \tau_\tau\,E_\tau}\,P_c\,\,\, ,
\label{eq:7}\\
\frac{dE_\tau}{dz}&=&- \left(\beta_\tau
+\gamma_\tau\,E_\tau\right) \,E_\tau \, \varrho_s \,\,\, .
\label{eq:8}
\end{eqnarray}
Here $m_\tau = 1.77$ GeV, $\tau_\tau \simeq 3.4 \times 10^{-13}\,$s
denotes the $\tau$ mean lifetime, whereas the parameters $\beta_\tau
\simeq 0.71 \times 10^{-6}$ cm$^2$ g$^{-1}$ and $\gamma_\tau \simeq
0.35 \times 10^{-18}$ cm$^2$ g$^{-1}$ GeV$^{-1}$, as discussed in
ref.\ \cite{Aramo:2004pr}, fairly describe the $\tau$ energy loss in
matter (for further references and a recent discussion, see
\cite{Dutta:2005yt}). We denote the transferred energy as $E^0_{\tau}
= E^0_{\tau}(E_\nu)=(1-\langle y_{CC}\rangle)E_\nu$ (see ref.\
\cite{Aramo:2004pr} for details). By solving Eq.s (\ref{eq:7}) and
(\ref{eq:8}) at the emerging point on the Earth surface one has
\begin{eqnarray}
P_c &=&  \left( F(E_\nu,E_\tau)\right)
^{\omega} \,
\exp\left\{-\frac{m_\tau}{c \tau_\tau \beta_\tau \varrho_s}
\left(\frac{1}{E_\tau}-
\frac{1}{E_\tau^0(E_\nu)}\right)\right\}\,\,\, , \label{eq:9}
\\
E_\tau & = & \frac{\beta_\tau \, E^0_{\tau}(E_\nu) \, \exp\left\{-
\varrho_s \, \beta_\tau (z_{\rm max}- z)\right\}} {\beta_\tau +
\gamma_\tau \, E^0_{\tau}(E_\nu) \left(1-\exp\left\{- \varrho_s \,
\beta_\tau (z_{\rm max}- z)\right\}\right)}\,\,\, , \label{eq:10}
\end{eqnarray}
where
\begin{equation}
F(E_\nu,E_\tau) \equiv
\frac{E^0_{\tau}(E_\nu)(\beta_\tau +
\gamma_\tau E_\tau) }{E_\tau(\beta_\tau +\gamma_\tau \, E^0_{\tau}(E_\nu)
)}\,\,\,,
\,\,\,\,\,\,\,\,\,
\omega \equiv \frac{m_\tau \, \gamma_\tau}
{c \tau_\tau \beta_\tau^2 \varrho_s}\,\,\,
.
\label{eq:10a}
\end{equation}
The quantity $z_{\rm max} = z_{\rm max}\left(\vec{r}_a\,,\Omega_a\right)$
represents the total length in rock for a given track entering the
lateral surface $\Sigma_a$ of the fiducial volume at the point
$\vec{r}_a$ and with direction $\Omega_a$.

The energy $E_\tau$ of the exiting lepton must be consistent with
Eq.\ (\ref{eq:10}). This condition is enforced by the presence of a
$\delta$-function in the final expression of the probability
$P_3$,
\begin{equation}
P_3=P_c \,\,\, \delta\left(E_\tau-\frac{\beta_\tau \,
E^0_{\tau}(E_\nu) \, \exp\left\{- \varrho_s \, \beta_\tau
(z_{\rm max}- z)\right\}} {\beta_\tau + \gamma_\tau \,
E^0_{\tau}(E_\nu) \left(1-\exp\left\{- \varrho_s \, \beta_\tau
(z_{\rm max}- z)\right\}\right)}\right)\,\,\, . \label{eq:11}
\end{equation}

\noindent 4) Once the $\tau$ lepton has emerged from the Earth
surface, its showering probability is determined by the total
distance it has to travel for reaching the fiducial volume. If we
denote with $\lambda_{\rm out} = \lambda_{\rm out} \left(\vec{r}_a\, ,
\Omega_a\right)$ the total length travelled in the atmosphere by
the $\tau$ before reaching the point $\vec{r}_a$ on the surface
$\Sigma_a$, and with $\lambda_{\rm in} = \lambda_{\rm in}
\left(\vec{r}_a\, , \Omega_a\right)$ the length of the
intersection of the track with the fiducial volume, the decay
probability inside the fiducial volume is given by
\begin{equation}
P_4 = \exp\left\{-\frac{\lambda_{\rm out} \, m_\tau} {c\tau_\tau \,
E_\tau}\right\} \, \left(1-\exp\left\{-\frac{\lambda_{\rm in} \,
m_\tau} {c\tau_\tau \, E_\tau}\right\}\right)\,\,\, .
\label{decay}
\end{equation}
Collecting together the different probabilities in Eq.s (\ref{eq:2}),
(\ref{eq:4}), (\ref{eq:11}) and (\ref{decay}), we have
\begin{equation}
k_a(E_\nu\,,E_\tau\,;\vec{r}_a\,,\Omega_a)=\int_0^{z_{\rm
max}}\,P_1 \, P_2 \, P_3 \, P_4 \,\,dz\,\,\, . \label{eq:12}
\end{equation}
Since the flux can be reasonably assumed isotropic, namely
\begin{equation}
\frac{d\Phi_\nu(E_\nu)}{dE_\nu\,d\Omega_a} =
\frac{d\Phi_\nu(E_\nu)}{dE_\nu\,d\Omega} \,\,\,\,\,\,\,\,\,\,
\forall a \,\,,
\end{equation}
we can rewrite Eq.\ (\ref{eq:1}), summing over all the surfaces, as
\begin{equation}
\frac{d N_\tau}{dt} = D \,\int dE_\nu \,
\,\frac{d\Phi_\nu(E_\nu)}{dE_\nu\,d\Omega} \,A(E_\nu) \,\,\, ,
\label{kernel1}
\end{equation}
where the effective aperture, $A(E_\nu) \equiv \sum_a A_a(E_\nu)$,
is the sum of each surface contribution,
\begin{equation}
A_a(E_\nu) \equiv \int dE_\tau\, K_a(E_\nu\,,E_\tau)\,\,\, ,
\label{aper}
\end{equation}
and
\begin{eqnarray}
K_a(E_\nu\,,E_\tau) &=& \int d\Omega_a \int  dS_a \,
\cos\left(\theta_a\right) \,\epsilon(E_\tau) \,
k_a(E_\nu\,,E_\tau\,;\vec{r}_a\,,\Omega_a) \nonumber \\ &=& \int
d\Omega_a \int  dS_a \, \cos\left(\theta_a\right)
\,\epsilon(E_\tau) \, \int_0^{z_{\rm max}}\,P_1 \, P_2 \, P_3 \,
P_4 \,\,dz\,\,\, . \label{kernel2}
\end{eqnarray}

In order to get the explicit expression for $K_a(E_\nu\,,E_\tau)$
an extremely involved integration has to be performed, which
requires the computation of all the properties for each track
($\vec{r}_a$, $\Omega_a$), taking into account the DEM of the
Auger site.  To this aim a suitable approach is based on the
following procedure: we use the available DEM of the Auger area to
isotropically generate a large number of oriented tracks (let us
say $N$) which cross the Auger fiducial volume. If we denote with
$N_a$ the subset of the $N$ tracks which enter through surface
$\Sigma_a$, then the kernel $K_a(E_\nu\,,E_\tau)$ can be well
approximated by the expression
\begin{equation}
K_a(E_\nu\,,E_\tau) \approx 2 \pi \, \epsilon(E_\tau) \,
\frac{S_a}{N_a} \, \sum_{i_a=1}^{N_a}
\cos\left(\theta_{i_a}\right) \,
k_a(E_\nu\,,E_\tau\,;\vec{r}_{i_a}\,,\Omega_{i_a})\,\,\, .
\label{kernel3}
\end{equation}

\begin{figure}[t]
\begin{center}
\epsfig{figure=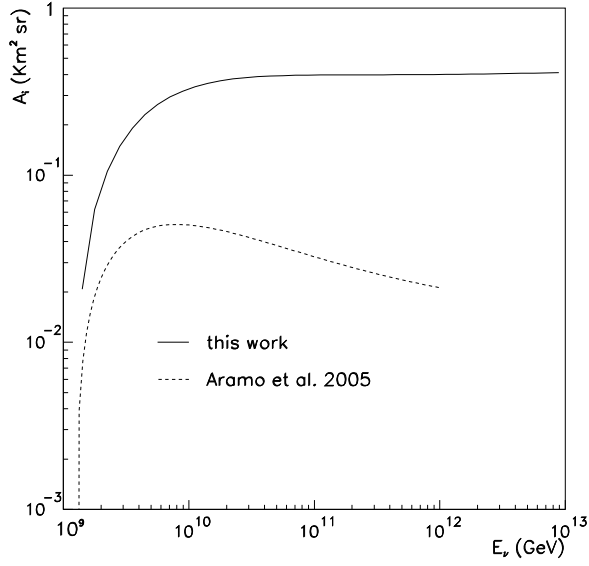,height=8cm} \caption{The total
effective aperture $A(E_\nu)$ is plotted versus the neutrino
energy (solid line). The dashed line corresponds to the same
quantity as obtained in ref.\ \cite{Aramo:2004pr} for $H=30$ km.}
\label{aperture}
\end{center}
\end{figure}
\begin{figure}[t]
\begin{center}
\epsfig{figure=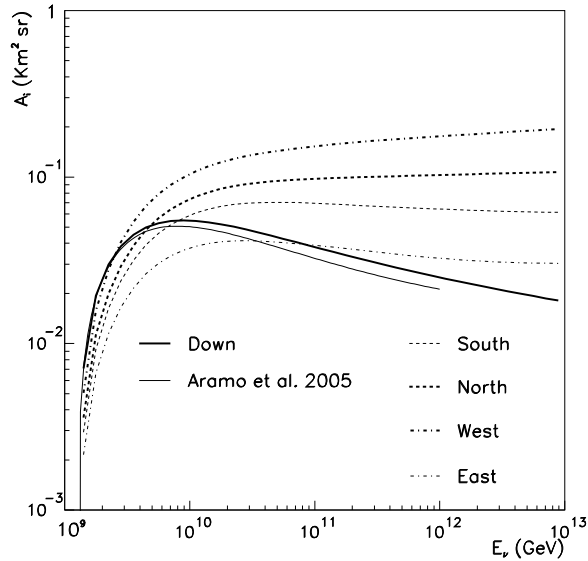,height=8cm} \caption{The effective
apertures $A_a(E_\nu)$ defined in Eq.\ (\ref{aper}) are plotted
versus the neutrino energy. The thin solid line corresponds to the
same quantity as obtained in ref.\ \cite{Aramo:2004pr} for $H=30$
km.} \label{apertures}
\end{center}
\end{figure}

We show in Fig.\ \ref{aperture} the aperture $A(E_\nu)$ as a
function of the neutrino energy (solid line), compared with the
results of ref.\ \cite{Aramo:2004pr} (dashed line). Remarkably, an
enhancement factor approximately of one order of magnitude is
found with respect to the semi-analytical results of ref.\
\cite{Aramo:2004pr}. The origin of the difference between the two
calculations can be explained in the following way. In ref.\
\cite{Aramo:2004pr} an event is rejected if the $\tau$ takes more
than 30 km to decay; this means that almost horizontal events
(where the $\tau$ emerges from the Earth far from the lower
surface of the experiment, travels more than 30 km before entering
in the fiducial volume, and then decays inside it), preferred by
the lateral surfaces due to the $\cos\left(\theta_{i_a}\right)$
term in Eq.\ (\ref{kernel3}), were not included in the calculation
of ref.\ \cite{Aramo:2004pr}, while they are present in this
analysis. In Fig.\ \ref{apertures} the different contributions
from each surface are plotted together with the results of ref.\
\cite{Aramo:2004pr} (thin solid line). As one can notice, the
contribution coming from the tracks going through the Down surface
(thick solid line) is in nice agreement with our previous
calculation. Moreover, it is possible to distinguish a North-South
effect and, more consistent, a West-East effect. 

It is worth observing the different high energy behavior of the
aperture corresponding to the D surface, $A_D$, with respect to
the others. The increase of the neutrino energy would select the
almost horizontal tracks, which however are depressed, for the
D-surface only with respect to the lateral ones, by the factor
$\cos\left(\theta_{i_D}\right)$.

\begin{table}[t]
\centering
\begin{tabular}{|c|c|c|c|c|c|}
\hline
& GZK-WB & GZK-L & GZK-H & NH & TD\\
\hline
Surface D & 0.016 & 0.040 & 0.095 & 0.246 & 0.100 \\
Surface S & 0.012 & 0.037 & 0.098 & 0.214 & 0.094 \\
Surface N & 0.015 & 0.046 & 0.125 & 0.267 & 0.120 \\
Surface W & 0.022 & 0.066 & 0.181 & 0.380 & 0.174 \\
Surface E & 0.008 & 0.024 & 0.061 & 0.139 & 0.060 \\ \hline
Total     & 0.074 & 0.213 & 0.560 & 1.245 & 0.548 \\ \hline\hline
Ref.\ \cite{Aramo:2004pr} & 0.02 & 0.04 & 0.09 & 0.25 & 0.11 \\
\hline
\end{tabular}
\caption{Yearly rate of Earth-skimming events at the FD for the
different neutrino fluxes considered in ref.\ \cite{Aramo:2004pr}.
The number of $\tau$'s showering into the fiducial volume that enter
through each lateral surface are reported, as well as the total number
of events for each flux. For comparison, we include the corresponding
results from ref.\ \cite{Aramo:2004pr}.}
\label{table::events}
\end{table}

We can use the expression in Eq.\ (\ref{kernel1}) to obtain the yearly
number of $\tau$ showering events at the FD of Auger (assuming a duty
cycle $D=10\%$). In Table \ref{table::events} these rates are reported
for the same UHE neutrino fluxes considered in section 2 of ref.\
\cite{Aramo:2004pr} (see in particular figs.\ 1 and 2), and described
in a series of papers
\cite{Waxman:1998yy,Kalashev:1999ma,Kalashev:2000af,Kalashev:2002kx,Semikoz:2003wv,Kachelriess:2003yy,Bhattacharjee:1998qc}.
The three GZK fluxes refer to three possible scenarios for cosmogenic
neutrinos, which are those produced from an initial flux of UHE
protons. Instead the NH (New Hadrons) and TD (Topological Defects)
cases are two examples of exotic models capable of generating the
UHECR above $10^{10}$ eV, with large associated neutrino fluxes.
For each neutrino flux (fixed column), we list the total number of
yearly events as well as the contributions from each lateral surface.

Some comments are in turn. The number of events crossing the
bottom surface is in fair agreement with the previous analytical
result of ref.\ \cite{Aramo:2004pr}. However, the total number of
events is a factor 4-6 larger (depending on the model considered),
showing that the main contribution to the number of events is
coming from almost horizontal showers, where the $\tau$ emerges
from Earth surface far away from the Auger fiducial volume and
decays inside it.  The enhancement factor depends on the different
features of the fluxes used in the analysis. For example, for the
three GZK models in Table \ref{table::events}, this factor ranges
from 3.7 to 6.2, corresponding to a hardening of the differential
fluxes in energy (see Figs.\ 1 and 2 of ref.\
\cite{Aramo:2004pr}). Instead, the enhancement is roughly the same
for the last two models in Table \ref{table::events} despite the
fact that they have a different spectrum in the high energy range,
due to the suppression of the very high energy neutrino events
which escape without showering.

As one can see from Table \ref{table::events}, a significant
difference in the number of events exists between the Surfaces W
(facing the Andes) and E, which shows a {\it mountain} effect.
This enhancement is mainly due to the largest amount of rock
encountered by horizontal tracks coming from the west side. If we
define, as a measure of the effect of mountains, the difference of
the number of expected events entering the volume from W-surface
and E-surface, divided by their average, the effect can be quite
remarkable (even order 30\%). This effect would also be larger by
comparing the exclusive apertures, but in this case the difference
in the apertures, which is larger for larger neutrino energy,
should compel the fast decreasing with energy of neutrino fluxes.
Of course, on the total number of events this 30\% effect is
diluted because it concerns only one surface among four lateral
ones. Moreover, 
a similar but smaller effect (of order $\sim$ 20-25\%) is also present
in the difference between the number of events from the Surfaces N
(facing the higher part of the {\it cordillera}) and S.

In conclusion, the largest contribution through the surfaces E, W,
S and N makes the total number of yearly Earth-skimming events
larger than the previous estimates of ref.\ \cite{Aramo:2004pr}.
This in turn increases the detection chances of UHE $\nu_\tau$ at
the FD of Auger, which seem realistic even for conservative neutrino
fluxes like GZK-WB, considering that data will be taken over many
years of observation.

\begin{table}[t]
\centering
\begin{tabular}{|c|c|c|c|c|c|}
\hline
& GZK-WB & GZK-L & GZK-H & NH & TD\\
\hline
$\nu_e$ & 0.034 & 0.098 & 0.277 & 0.565 & 0.276 \\
$\nu_\tau$ & 0.009 & 0.018 & 0.036 & 0.113 & 0.043 \\\hline
Total & 0.043 & 0.116 & 0.313 & 0.678 & 0.319 \\ \hline
\end{tabular}
\caption{Yearly expected number of down-going events at the FD, due to
the showering of $\nu_e$ and $\nu_\tau$ inside the Auger fiducial
volume. The different predictions refer to the same fluxes of Table
\ref{table::events} and to a zenith angle larger than $60^\circ$.}
\label{table::number}
\end{table}
In analogy with the analysis of ref.\ \cite{Aramo:2004pr}, we report
in Table \ref{table::number} the expected numbers of yearly down-going
events at FD, produced by UHE $\nu_e$ and $\nu_\tau$ showering inside
the fiducial volume\footnote{From the study of
$\nu_e$,$\nu_\mu$--induced shower in the atmosphere of ref.\
\cite{Ambrosio:2003nr} it is argued that high energy $\nu_\mu$'s have
a very small probability of inducing a shower before reaching the
ground.}. The numbers are calculated in a similar way to the
Earth-skimming case, but with a different kernel in Eq.\
(\ref{eq:12}),
\begin{equation}
k_a (E_{\nu_\alpha}\,, E_\alpha\, ; \vec{r}_a\,, \Omega_a) =
\delta \left( E_\alpha - E^0_\alpha (E_{\nu_\alpha}) \right)~
P_1^{\nu_\alpha}~P_2^{\nu_\alpha}
\end{equation}
where $\alpha=e,\tau$ and $E^0_\alpha (E_{\nu_\alpha})$ is the
transferred energy to the charged lepton. For a $\nu_e$ we have
\begin{eqnarray}
P_1^{\nu_e}&=&\exp
\left\{ - \frac{\lambda_{\rm out}}{\lambda_{CC}^\nu (E_{\nu_e})}
\right\} \nonumber \\
P_2^{\nu_e}&=&
\int_0^{\lambda_{\rm in}}\, dz~ \frac{\exp\left\{- z /
\lambda_{CC}^\nu (E_{\nu_e})\right\}}{\lambda_{CC}^\nu
(E_{\nu_e})}=
1 - \exp \left\{ - \frac{\lambda_{\rm in}}{\lambda_{CC}^\nu
(E_{\nu_e})} \right\} \,\,\,,
\end{eqnarray}
which take into account the probability that the neutrino survives
until the fiducial volume producing an electron which initiates a
shower inside the detector. Instead, for a $\nu_\tau$ we have
\begin{eqnarray}
P_1^{\nu_\tau}&=& 1 - \exp \left\{ - \frac{\lambda_{\rm in}\,
m_\tau}{c\tau_\tau \, E_\tau} \right\} \nonumber \\
P_2^{\nu_\tau}&=&
\int_0^{\lambda_{\rm out}}\, dz~ \frac{\exp\left\{-
z / \lambda_{CC}^\nu (E_{\nu_\tau})\right\}~
\exp\left\{- (\lambda_{\rm out}-z)\,
m_\tau/(c\tau_\tau \, E_\tau)\right\} }
{\lambda_{CC}^\nu (E_{\nu_\tau})} \nonumber \\
&=&\frac{ \exp\left\{-
\lambda_{\rm out}/\lambda_{CC}^\nu (E_\nu)\right\} -
\exp\left\{- \lambda_{\rm out}\,
m_\tau /(c\tau_\tau \, E_\tau)\right\} }{ [\lambda_{CC}^\nu (E_\nu)\,
m_\tau/(c\tau_\tau \, E_\tau)] - 1}\,\,\,,
\end{eqnarray}
which take into account the probability that the neutrino produces a
$\tau$ which survives until the fiducial volume and initiates a shower
inside the detector.

In this letter a new and more detailed computation of the aperture for
Earth-skimming $\nu_\tau$ of the fluorescence detector at the Pierre
Auger Observatory has been performed. The evaluation has been carried
out by using a different approach with respect to the
previous semi-analytical one of ref.\ \cite{Aramo:2004pr} and
taking into account the real elevation profile of the area around
Auger. The obtained results show an increase of the effective aperture, and
correspondingly of the number of the expected 
events. This larger result is mainly due to the 
contribution of almost horizontal Earth-skimming $\tau$'s, emerging
from the Earth surface far away from Auger but decaying inside its
fiducial volume. Remarkably the presence of mountains near Auger leads
to an enhancement factor on the total number of events of the order
1.3-1.6. As previously shown (see e.g.\ the discussion in
\cite{Zas:2005zz}), our results indicate that the number of UHE neutrino
events at Auger is comparable for showers induced by down-going UHE
neutrinos and by the decays of $\tau$'s produced by Earth-skimming UHE
$\nu_\tau$'s.

It is worth observing that the efficiency of the FD detector,
whose parameterization we have used in Eq.\ (\ref{eq:1}), can be
considered as oversimplified and that one should expect a
behaviour of FD much more complicated with energy and/or geometry
of the event. Unfortunately, a complete and exhaustive analysis of
this issue is still absent and thus not available for any analysis
like ours. Nevertheless, our predictions can be considered as a
first but probably useful step in the right direction, which
should motivate more study inside the Auger collaboration for
example in order to obtain a reliable efficiency of FD with energy
and for horizonal showers.

Possible future improvements of the present work should take properly
into account the response of the flourescence detector, as well as the
real geometry of the Auger area and the fact that some of the UHE
$\nu_\tau/\tau$ tracks could partly traverse water.

\section*{Acknowledgments}

We thank M.\ Ambrosio, F.\ Guarino, L.\ Perrone and R.\ V\'azquez for
useful discussions.  This work was supported by a Spanish-Italian AI,
the Spanish grants BFM2002-00345 and GV/05/017 of Generalitat
Valenciana, as well as a MEC-INFN agreement. SP was supported by a
Ram\'{o}n y Cajal contract of MEC.

\end{document}